\documentclass[aps,prl,showpacs,showkeys,twocolumn,10pt]{revtex4-1}
\usepackage{graphicx,amsmath,amsfonts,amssymb,dsfont}
\usepackage[activeacute,english]{babel}

\usepackage{graphicx}
\usepackage{amssymb}
\begin{document}
\title{Stochastic thermodynamics for kinetic equations.}
\author{C. Van den Broeck}
\affiliation{Hasselt University, B-3590 Diepenbeek, Belgium}
\author{R. Toral}
\affiliation{ IFISC (Instituto de F{\'\i}sica Interdisciplinar y Sistemas Complejos), Campus UIB, Palma de Mallorca, Spain}
\date{\today}
\begin{abstract}
Stochastic thermodynamics is formulated for variables that are odd under time reversal. The invariance under spatial rotation of the collision rates due to the isotropy of the heat bath is shown to be a crucial ingredient. An alternative detailed fluctuation theorem is derived, expressed solely in terms of forward statistics. It is illustrated for a linear kinetic equation with kangaroo rates.
\end{abstract}
\maketitle
The second law of thermodynamics is arguably one of the most general laws of nature. While originally stipulating the increase of total entropy in a closed isolated system $\Delta S_{tot}\geq 0$, it was reformulated by splitting the entropy change $\Delta S$ of an open system into the sum $\Delta S=\Delta_i S +\Delta_e S$ of a non-negative entropy production term $\Delta_i S\geq 0$ plus an entropy exchange contribution $\Delta_e S$. In particular when in contact with a single heat bath at temperature $T$, the exchange is given  by $\Delta_e S=Q/T$, where $Q$ is the  amount of heat into the system.  Over the past two decades, a much deeper formulation of the second law has been achieved by focusing on small open systems. One can still define all the above mentioned quantities, but they will now fluctuate  from one measurement to another. Using lower case to distinguish the values from the non-fluctuating macroscopic counterparts, one has $\Delta s=\Delta_i s +\Delta_e s$, with $\Delta_e s=q/T$. The second law  is replaced by a symmetry property for the probability density $P(\Delta_i s)$ to observe an entropy production $\Delta_i s$. In its simplest form, the so-called fluctuation theorem states that the probability for observing an entropy increase is exponentially larger than that for observing a corresponding decrease,
$P(\Delta_i s)=\exp(\Delta_i s){P}(-\Delta_i s)$. The second law $\langle \Delta_i s \rangle \geq 0$ follows as a subsidiary result. The fluctuation theorem has been obtained at different levels of description, ranging from the microscopic laws \cite{Jarzynski:1997,Kawai:2007}, over thermostated systems \cite{Evans:1993,Gallavotti:1996} to stochastic dynamics \cite{Crooks:1998,Kurchan:1998,Lebowitz:1999}. ``Stochastic thermodynamics" is easy to formulate in the context of a Markovian description, both at the level of a Langevin/Fokker-Planck equation or the more general Master equation \cite{Sekimoto:2010,Seifert:2012,VandenBroeck:2013,VandenBroeck:2014b}, and its predictions have by now been confirmed by numerous experiments. The focus has been mostly on overdamped systems with variables that are even under time-reversal. However, for variables, such as velocities instead of positions, it was claimed that the theory becomes more involved and hence loses some of its  appeal \cite{Spinney:2012,Lee:2013}. In this Letter, we show that this is not the case if the transition probabilities obey, in addition to detailed balance, a symmetry property, reflecting the isotropy of the heat bath. To demonstrate the role and importance of this condition, we develop the stochastic thermodynamics, both at the ensemble and trajectory level, for linear kinetic equations, a field that has not been explored before and for which there is a large potential interest. We derive the fluctuation theorem, including a new version expressed only in terms of probabilities computed from the forward process. As a application, we provide explicit illustrations for  the special case of kinetic ``kangaroo" equations \cite{VandenBroeck:2014}.

We consider the simplest scenario of a system consisting of a single stochastic Maxwell-Lorentz particle, cf. \cite{Gradenigo2012,Gradenigo2013} for a detailed analysis of a similar model,  with mass $m$, velocity $v$ at position $x$ in the constant external force field $F$ (acceleration $a=F/m$), and in contact with a single isotropic heat reservoir at rest with temperature $T$.  The stochastic dynamics of the particle is characterized by a probability density $P(x,v;t)$, obeying the linear kinetic equation:
\begin{widetext}
\begin{equation}\label{master_general}
\left[\frac{\partial \,}{\partial t}+v\frac{\partial \,}{\partial x}+a\frac{\partial \,}{\partial v}\right]P(x,v;t)=\int dv'\left[k(v'\to v)P(x,v';t)-k(v\to v')P(x,v;t)\right].
\end{equation}
\end{widetext}
Here $k(v'\to v)$  is the transition probability per unit time (rate) for a change of velocity from $v'$ to $v$. 
Formulation of the first law at the trajectory level is straightforward.  The energy $e(t)$ of a particle in the constant external force field $F$ is:
\begin{equation}
{e}(t)=-F  x(t)+\frac{1}{2}m v^2(t),
\end{equation}
where $x(t)$ and $v(t)$ are the position and velocity of the particle at time $t$ in the given realization. The ``ensemble" version of the first law is obtained by averaging with respect to the probability density $P(x,v;t)$:
\begin{eqnarray}
{E}(t)=\langle{e}(t)\rangle=-F \langle x(t)\rangle+\frac{1}{2}m\langle v^2(t) \rangle.
\end{eqnarray}
In-between collisions, potential energy is converted into kinetic energy following Newton's law $m\dot{v}(t)=F$, hence this non-dissipative process produces no net energy $\dot{e}(t)=0$, and neither  work nor heat are exchanged. The punctual collisions with the heat bath however lead to an instantaneous exchange of energy under the form of heat:
\begin{eqnarray}
\dot{e}(t)=\dot{q}(t)\label{heattraj},
\end{eqnarray}
with $\dot{q}(t)$ a sum of delta functions at the instants of the collision and with amplitude $\frac12m(v^2-{v'}^2)$ for a collision changing the velocity from $v'$ to $v$. At the ensemble level, the resulting heat flux $\dot{Q}(t)$ is obtained by averaging over the frequency of such collisions:
\begin{equation}
\dot{E}(t)=\dot{Q}(t)
= \int \hspace{-8pt}\int dvdv'k(v'\to v)P(v',t)\frac{1}{2}m(v^2-{v'}^2).\label{heat}
\end{equation}
 
We next turn to the second law and formulate it first at the ensemble level. The ``ensemble" entropy associated to the distribution $ P(x,v;t)$ is given by $S(t)=-k_B\int dxdv P(x,v;t)\ln P(x,v;t)$, with $k_B$ Boltzmann's constant. When considering the time derivative of this quantity, we note that the motion is purely Hamiltonian in-between collisions. Following  Liouville's theorem, this part of the dynamics leaves  the entropy invariant \cite{Balescu:1975}.  Hence, we need only to focus on the change of the entropy induced by the dissipative collisions, affecting solely the velocity variables. From: 
\begin{equation}
S(t)=-k_B\int dv\, P(v,t)\ln P(v,t)
\end{equation}
we find in combination with the evolution equation for $P(v,t)$, obtained from Eq.(\ref{master_general}), and following some simple manipulations, that the rate of change of the entropy is given by:
\begin{eqnarray}
\dot S
&=&k_B\hspace{-4pt} \int \hspace{-8pt}\int dv\,dv' k(v'\to v)P(v',t)\ln \frac{P(v',t)}{P(v,t)}.
\end{eqnarray}
This rate of entropy change can thus be rewritten under the standard form $\displaystyle\dot S=\dot S_i+\dot S_e$ 
 with the rates of ``entropy production" and ``entropy exchange" given by:
\begin{eqnarray}
\frac{\dot S_i}{k_B}&=& \hspace{-4pt}\int \hspace{-8pt}\int dvdv'k(v'\to v)P(v',t)\ln \frac{k(v'\to v)P(v',t)}{k(v\to v')P(v,t)}\ge 0,\nonumber\\
\frac{\dot S_e}{k_B}&=& \hspace{-4pt}\int \hspace{-8pt}\int dvdv'k(v'\to v)P(v',t)\ln \frac{k(v\to v')}{k(v'\to v)}.
\end{eqnarray}

These results are mathematically exact but, in order to achieve a correct thermodynamic interpretation of the entropy production and exchange, one  needs in addition proper physical  input about the collision mechanism, i.e. about the collision rate.
We focus here on the simplest case in which the collision process represents energy exchange with a single isotropic thermal reservoir at temperature $T$. As a result the collision process must induce, in absence of an external force, a relaxation to the Maxwell-Boltzmann  distribution $\varphi_0$, i.e., one has:
\begin{eqnarray}\label{max}
&&\int dv' k(v\to v')\varphi_0(v)=\int dv' k(v'\to v)\varphi_0(v'),\\
\label{phist}
&&{\textrm{with   }} \varphi_0(v)= \frac{e^{-v^2/2\sigma^2}}{\sigma\sqrt{2\pi}}\;\;\;\;\; \sigma^2=mk_BT.
\end{eqnarray} 
As was realised first by Onsager \cite{Balescu:1975}, micro-reversibility leads to a more stringent condition of detailed balance: 
\begin{eqnarray}\label{db}
k(v\to v')\varphi_0(v)= k(-v'\to -v)\varphi_0(-v').
\end{eqnarray}
This detailed balance relation involves velocity inversion, and seems to be at variance with the condition Eq.~(\ref{max}).
The discrepancy is solved by making the crucial observation that, for a collision describing heat exchange with an isotropic bath, there is an additional symmetry requirement of invariance under reflection (and more generally under rotation \cite{Gaspard:2013,gaveau}):
\begin{eqnarray}\label{rs}
k(v'\to v)= k(-v'\to -v).
\end{eqnarray}
With this extra condition, the detailed balance relation Eq.~(\ref{db}) implies Eq.~(\ref{max}).

Eq.~(\ref{db}) allows to make the consistent connection between first and second laws: the entropy exchange $\dot S_e$ can be rewritten ($\varphi_0(-v)=\varphi_0(v)$):
\begin{eqnarray}
&&\dot S_e= k_B\hspace{-4pt}\int \hspace{-8pt}\int dvdv'k(v'\to v)P(v',t)\ln \frac{\varphi_0(v')}{\varphi_0(v)}=\frac{\dot Q}{T},\hspace{30pt} \label{see}
\end{eqnarray}
%\begin{eqnarray}
%&&\dot S_e= k_B\hspace{-4pt}\int \hspace{-8pt}\int dvdv'k(v'\to v)P(v',t)\ln \frac{k(-v\to -v')}{k(v'\to v)}\nonumber\\
%&&= k_B\hspace{-4pt}\int \hspace{-8pt}\int dvdv'k(v'\to v)P(v',t)\ln \frac{\varphi_0(v')}{\varphi_0(v)}=\frac{\dot Q}{T},\label{see}
%\end{eqnarray}
where ${\dot Q}$ is the rate of energy (heat) exchange from the bath to the particle, cf. Eqs.(\ref{heat},\ref{phist}).
The entropy production is zero if and only if $k(v'\to v)P(v')= k(v\to v')P(v)$, implying that $P(v)/P(v')=\varphi_0(v)/\varphi_0(v')$ and hence $P(v)=\varphi_0(v)$. We conclude that entropy production vanishes if and only if  detailed balance is satisfied.

We now show that both Eq. (\ref{db}) and Eq. (\ref{rs}) are crucial to formulate the second law at the trajectory level.  
The stochastic entropy for the velocity variables reads \cite{Seifert:2012}:
\begin{equation}\label{defs}
s(t)=-k_B \ln P(v(t),t).
\end{equation}
Note that this entropy still retains an ensemble character, as one needs to specify the probability distribution $P(v,t)$, which is the probability to observe the particle with velocity $v$ at time $t$ starting from some specific initial probability distribution.  This so-called forward experiment is ran from initial time $t_i$ to some final time $t_f$. We now write:
\begin{equation}\label{sb}
{\dot s}={\dot s_i}+{\dot s_e},
\end{equation}
where the trajectory entropy exchange is the obvious analogue of the ensemble value given in Eq.~(\ref{see}): ${\dot s_e}={\dot q}/{T}$.
The meaning of the trajectory entropy production is most easily clarified by integrating Eq.~(\ref{sb}) over a finite time, leading to the finite difference balance:
\begin{equation}
\Delta s=\Delta_i s+\Delta_e s,
\end{equation}
with $\Delta_e s=\displaystyle q/T$ and $q$ is the total amount of heat received (by collisions) from the heat bath in the  realization under consideration. 
An elegant derivation of the celebrated fluctuation theorem for the trajectory entropy production proceeds with the consideration of the probability for a trajectory in forward and reverse dynamics. We consider the simplest case of  steady state operation, with the initial state of the forward experiment under acceleration $a$ sampled from the steady state distribution $P^{st}_a(v)$. The reverse trajectory proceeds under the same acceleration $a$, starting with the final distribution of the forward probability, but with inverted speeds. Its properties will be identified with a superscript tilde.  Let $P(\Pi)$ and $\tilde{P}(\tilde{\Pi})$ denote the probabilities  for a forward and reverse trajectory, $\Pi$ and $\tilde{\Pi}$,  
%in the forward and reverse dynamics, 
respectively.  One now verifies the following striking equality:
\begin{equation}\label{dft}
\Delta_i s=k_B\ln \frac{P(\Pi)} {\tilde{P}(\tilde{\Pi})}.
\end{equation}
The proof goes as follows. The probability of a trajectory involves the initial probability, the probability for not having collisions in-between the transitions, and the probability for transitions. Since the starting probability of the reverse dynamics is equal to the final probability of the direct dynamics, the log ratio of the initial probability contributions reproduces $\Delta s=k_B \ln P(v(t_f),t_f)-k_B \ln P(v(t_i),t_i)$, cf. Eq.~(\ref{defs}). Due to the detailed balance condition Eq.~(\ref{db}), the log ratio of probabilities for collisions in forward and backward dynamics, cf. $\ln k(v'\to v)/k(-v\to -v')=\ln \varphi_0(v)/\varphi_0(v')= m (v^2-v'^2)/(2 k_B T)$, reproduces $-\Delta_e s=-q/T$. Finally, due to the reflection symmetry Eq.~(\ref{rs}), the probability for having no collisions, determined by the rates $k(v'\to v)$ and $k(-v'\to -v)$ when we have a velocity $v'$ and $-v'$, respectively, is the same in forward and backward trajectories. Hence the corresponding terms cancel out, and we have $\Delta_i s=\Delta s-\Delta_e s$ as required. We conclude that both at the ensemble level and at the trajectory level, the combination of detailed balance condition with the reflection symmetry are essential for a consistent stochastic thermodynamic interpretation. 
The implications of Eq.~(\ref{dft}) are well known  \cite{Esposito:2010}: the probability distributions $P(\Delta_i s)$ and $\tilde{P}(-\Delta_i s)$ for observing an entropy production $\Delta_i s$ in the forward process and minus this value in the backward process  obey a detailed fluctuation theorem:
\begin{equation}\label{dfth}
 \frac{P(\Delta_i s)} {\tilde{P}(-\Delta_i s)}=\exp(\Delta_i s),
\end{equation}
from which follows the integral fluctuation theorem: $\langle \exp(-\Delta_i s)\rangle=1$.
A comment concerning the interpretation of Eq.~(\ref{dfth}) is in place, for more details see \cite{VandenBroeck:2013,spinney:2013,VandenBroeck:2014b,Becker:2014,Harris:2007}.  In general $-\Delta_i s$ is not the entropy production of  the reverse trajectory. This will only be the case if the inverse ``tilde" process is an involution, i.e., twice this operation is equal to the identity. In particular, the final probability distribution of the reverse process should be equal to the initial distribution of the forward process. In the case of even variables, a sufficient condition is that the forward  process starts and ends in a steady state. For odd variables, this condition is not sufficient as is illustrated by the above example: the velocity inversion at the end of the forward process produces a probability distribution that is no longer at the steady state when $a \neq 0$. There is however a simple procedure to cure this problem and to obtain a detailed fluctuation theorem which is, just like the integral fluctuation theorem, expressed solely in terms of a (slightly modified) forward process. At the end of the forward process, one  performs an instantaneous switch of the probability distribution from $P(v_f)$ to $P(-v_f)$, implying and entropy change of  $\Delta_{vi} s=\ln P(v_f)/P(-v_f)$. This is, on average (with respect to $P(v_f)$), an irreversible entropy producing step. With this additional step, velocity inversion at the end of the forward will  reproduce the steady state distribution, which is also in the case considered here the initial distribution  of the forward process. In conclusion the corrected entropy production $\Delta_i s_c=\Delta_i s+\Delta_{vi} s$ will obey a symmetric detailed fluctuation theorem:
\begin{equation}\label{dfthc}
 \frac{P(\Delta_i s_c)} {{P}(-\Delta_i s_c)}=\exp(\Delta_i s_c),
\end{equation} 
which  can conveniently be verified by considering statistics of the forward experiment alone.
\begin{figure}
\centerline{\includegraphics[width=1.\linewidth]{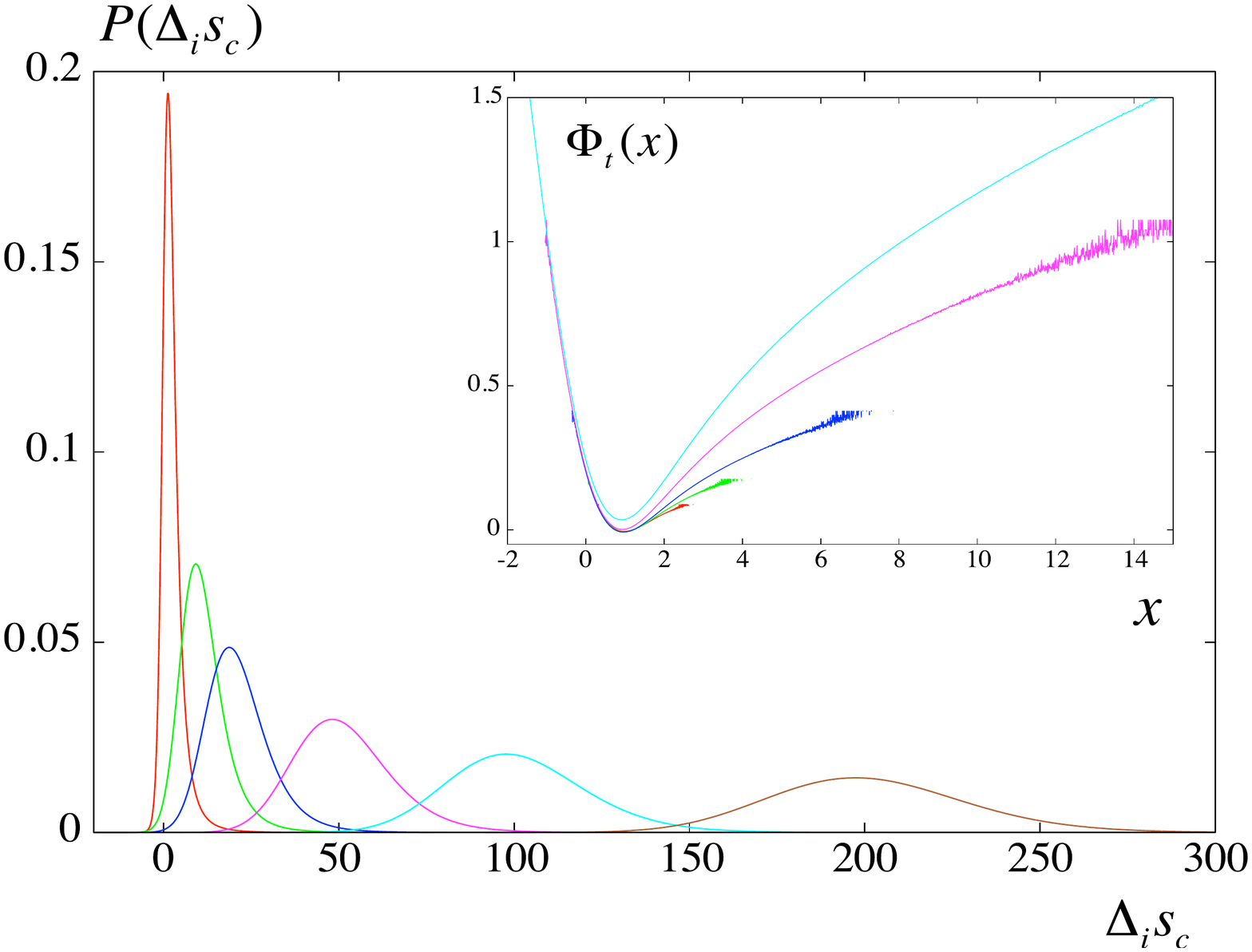}}
\vspace{-0.5cm}\caption{Probability distribution $P(\Delta_i s_c)$ (main plot) and large deviation function $\Phi_t(x)$ (inset) obtained from a numerical simulation  of the stochastic process Eq.~(\ref{master_general}) for the case of a kangaroo reaction rate (\ref{rate1}) with a uniform rate $\lambda(v)=1/\tau$, and operating under steady state conditions. We have taken $\tau=1$, the acceleration $a=1$ and $\sigma=1$ in (\ref{phist}). From left to right in the main plot the curves correspond to $t=1,10,20,50,100,200$. In the inset we see that $\Phi_t(x)$ converges for large time to a time-independent curve, the large deviation function. The histograms have been obtained after averaging for $4\times10^{11}$ realizations. \label{fig:uni}}
\end{figure}

\begin{figure}
\centerline{\includegraphics[width=1.\linewidth]{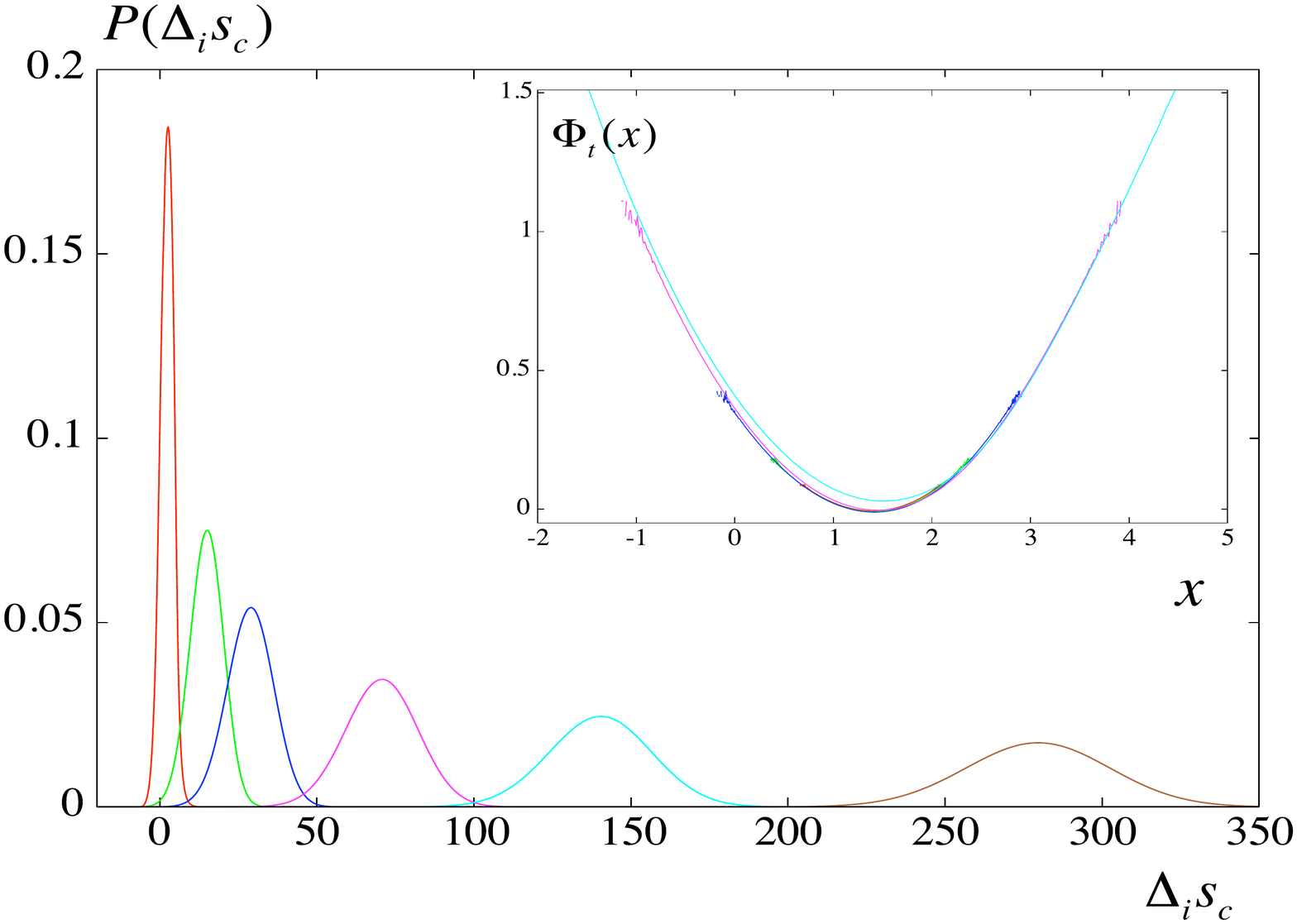}}
\vspace{-0.5cm}\caption{Similar to Fig.~\ref{fig:uni} for the case of a rate proportional to the absolute value of the velocity $\lambda(v)=\alpha |v|$ with $\alpha=1$. Same parameter values and time sequence as in Fig.~\ref{fig:uni}.\label{fig:absv}}
\end{figure}

\begin{figure}
\centerline{\includegraphics[width=1.\linewidth]{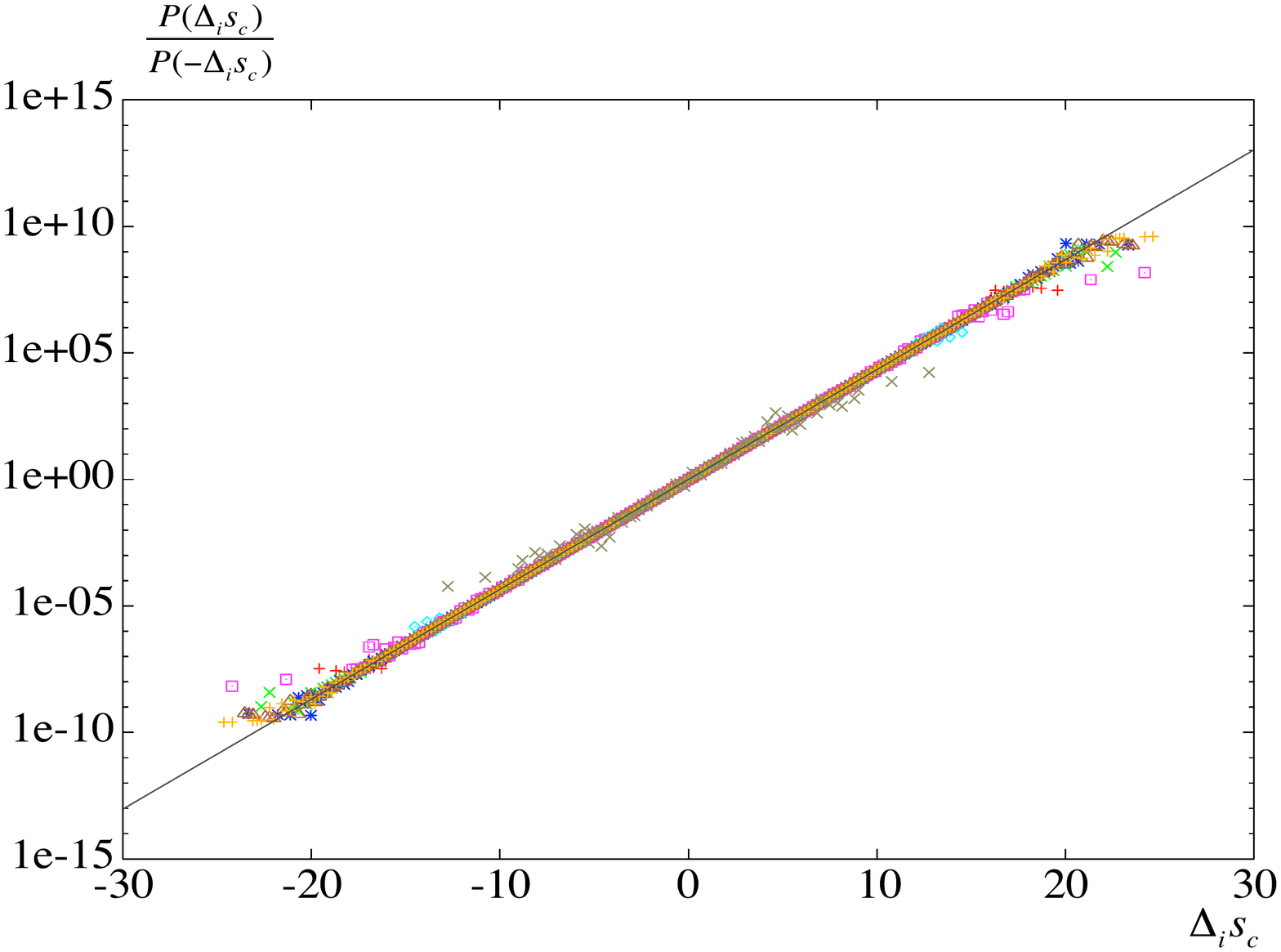}}
\vspace{-0.5cm}\caption{Test of the fluctuation theorem including the data from the histograms of Figs.~\ref{fig:uni} and \ref{fig:absv}.\label{fig:ft}}
\end{figure}

To illustrate the above formalism, we focus on the simple case of a ``kangaroo" kinetic equation with a rate $k(v'\to v)$ \cite{VandenBroeck:2014}: 
\begin{equation}\label{rate1}
k(v'\to v)=\lambda(v')\varphi(v).
\end{equation}
One  verifies that the detailed balance symmetry Eq.~(\ref{db}) implies in this case  that the collision rate $\lambda=1/\tau$  is a constant, independent of $v'$, and hence $k(v'\to v)={\varphi_0(v)}/{\tau}$. 
The reflection symmetry Eq.~(\ref{rs}) is, in this case, an automatic consequence of the detailed balance condition Eq.~(\ref{db}). Numerical simulations of the stochastic process Eq.~(\ref{master_general}) allow us to compute the probability distribution $P(\Delta_is_c)$, see Fig.~\ref{fig:uni}, and test the validity of the fluctuation theorem, cf. Fig. \ref{fig:ft}. In the inset of Fig.~\ref{fig:uni} we plot the large deviation function $\Phi_t(x)$ that results of the fit $P(\delta_i s_c)\sim \exp\left[-t\Phi_t(\delta_i s_c)\right]$, with $t=t_f-t_i$ and $\delta_i s_c=\frac{\Delta_i s_c}{t}$. We  have also considered the case $\lambda(v)=\alpha |v|$ whose corresponding results for  $P(\Delta_is_c)$ are shown in Fig.~\ref{fig:absv}. Interestingly, the reflection symmetry property is still satisfied, and the detailed fluctuation theorem is formally recovered. However the detailed balance condition is violated.  The steady state solution is not Maxwellian, and the interpretation of $\Delta_is_c$ as thermodynamic entropy production is false.

We close with a few remarks. Stochastic thermodynamics has been developed in great detail for Langevin equations, see e.g. \cite{Seifert:2012,Tome2015}, both in the over-damped and underdamped. A well documented case is a chain of harmonic oscillators in contact with two heat baths, see \cite{Saito:2007,Saito:2011,Kundu:2011,Fogedby:2012}. One may wonder why the symmetry property Eq.~(\ref{db}) has not been discussed in this context. By making the diffusion approximation on the master equation (\ref{master_general}) \cite{vanKampen:2007}, one easily verifies that Eq.~(\ref{db}) requires that the drift term be uneven in the velocity and the noise term even. These conditions are met in a generic Langevin equation, explaining why this issue has not appeared in this context. The formalism presented above can be easily extended to more complicated situations, such as multiple particles with  vectorial velocities in contact with several reservoirs of heat, particles or momentum and with time-dependent external forcing. Also the splitting of the entropy production in several components, such as the adiabatic and non-adiabatic contribution, proceeds as before \cite{Esposito:2010}.

We acknowledge financial support from EU (FEDER) and the Spanish MINECO under Grant INTENSE@COSYP (FIS2012-30634) and the MO 1209 COST action.

%\bibliography{References_of_my_papers-Kangaroo_kinetic_equation}
\bibliography{kinetic_equation_v9}

\end{document}